\documentclass[preprint,aps,nofootinbib]{revtex4}
\pdfoutput=1
\usepackage{graphicx}
\usepackage{epstopdf}    
\usepackage{dcolumn}
\usepackage{amsfonts}
\usepackage{amsmath}
\usepackage{amssymb}
\usepackage{xcolor}
\usepackage{bm}
\usepackage{color}
\usepackage{comment}
\usepackage{epsf}
\usepackage{float}
\usepackage{enumitem}

\usepackage[colorlinks=true,citecolor=blue]{hyperref}
\hypersetup{colorlinks=true,citecolor=blue,linkcolor=red,urlcolor=blue}

\counterwithin{equation}{section}

\newcolumntype{L}{>{\centering\arraybackslash}m{3cm}}
	
	\def\ast{{\dagger}}

	\newcommand{\bqa}{\begin{eqnarray}}
	\newcommand{\eqa}{\end{eqnarray}}
	\newcommand{\bwt}{\begin{widetext}}
	\newcommand{\ewt}{\end{widetext}}

	\newcommand{\nn}{\nonumber \\}

	\newcommand{\del}{\delta}

	\newcommand{\ti}{\tilde}

\newcommand{\edc}{\end{document}}
\newcommand{\bb} {\begin{thebibliography}}
\newcommand{\eb}{\end{thebibliography}}
\newcommand{\bi}[1]{\bibitem{#1}}
\newcommand{\bc}{\begin{center}}
\newcommand{\ec}{\end{center}}

\newcommand{\be}{\begin{equation}}
\newcommand{\ee}{\end{equation}\normalsize}
\newcommand{\bea}{\begin{eqnarray}}
\newcommand{\eea}{\end{eqnarray}}
\newcommand{\ba}{\begin{array}{l}   }

\newcommand{\ea}{\end{array}}

\newcommand{\ci}{\cite}

\def\bfr{{\bf r}}
\def\bfk{{\bf k}}

\newcommand{{\vergul}}{  ,}

\newcommand{\veps}{\varepsilon }

\renewcommand{\theequation}{\arabic{section}.\arabic{equation}}

\begin{document}

\title{Defining a critical temperature of a crossover from BEC to the normal phase in anisotropic quantum magnets}

	\author{Abdulla Rakhimov$^1$, Asliddin Khudoyberdiev$^1$, Zabardast Narzikulov$^1$, Bilal Tanatar$^2$ }
	\affiliation{$^1$Institute of Nuclear Physics, Tashkent 100214, Uzbekistan \\
	$^2$Department of Physics, Bilkent University, Bilkent, 06800 Ankara, Turkey}

		\date{\today}
		
		\begin{abstract}
We address the problem of identifying the critical temperature in a crossover from the Bose-Einstein condensed (BEC) phase to the normal phase. For this purpose we study the temperature dependence of magnetization of spin-gapped quantum magnets described by BEC of triplons. We have calculated the heat capacity $C_H$ at constant field and fluctuations in magnetization in a spin-gapped quantum magnet using the Hartree-Fock-Bogouliubov approximation and found optimized parameters of the Hamiltonian of triplon gas. In the region of phase transition, the heat capacity $C_H$ is
smeared out due to the Dzyaloshinsky-Moriya (DM) interaction. The sharp maximum
of the fluctuations in the magnetization is identified as the critical temperature of the crossover.
  		\end{abstract}

\maketitle

\section{Introduction}\label{sec1}

\begin{figure}[h]%
\begin{minipage}[H]{0.4\linewidth}
\hfill
{\includegraphics[width=1.4\textwidth]{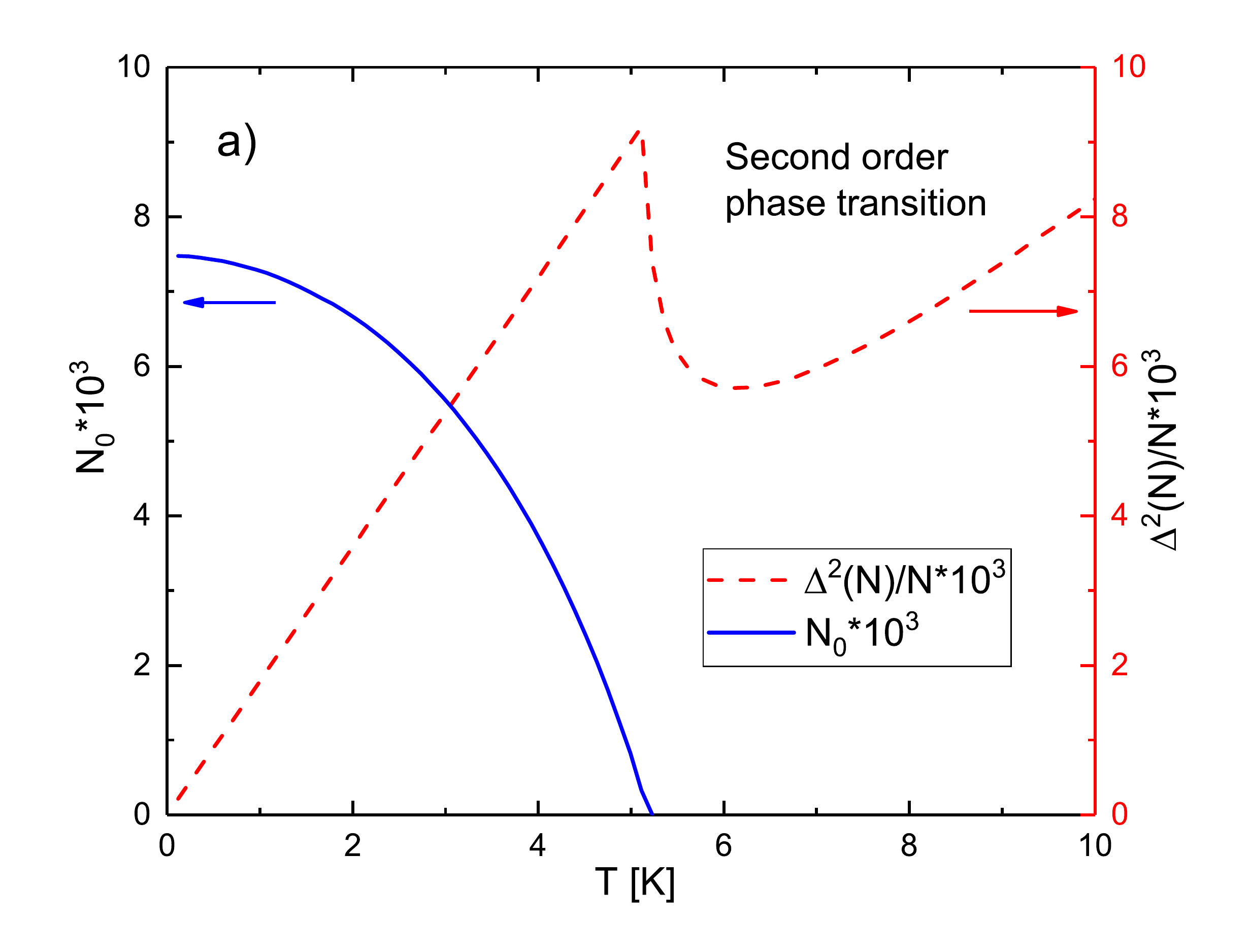}}
\end{minipage}
\qquad\qquad
\begin{minipage}[H]{0.4\linewidth}
{\includegraphics[width=1.4\textwidth]{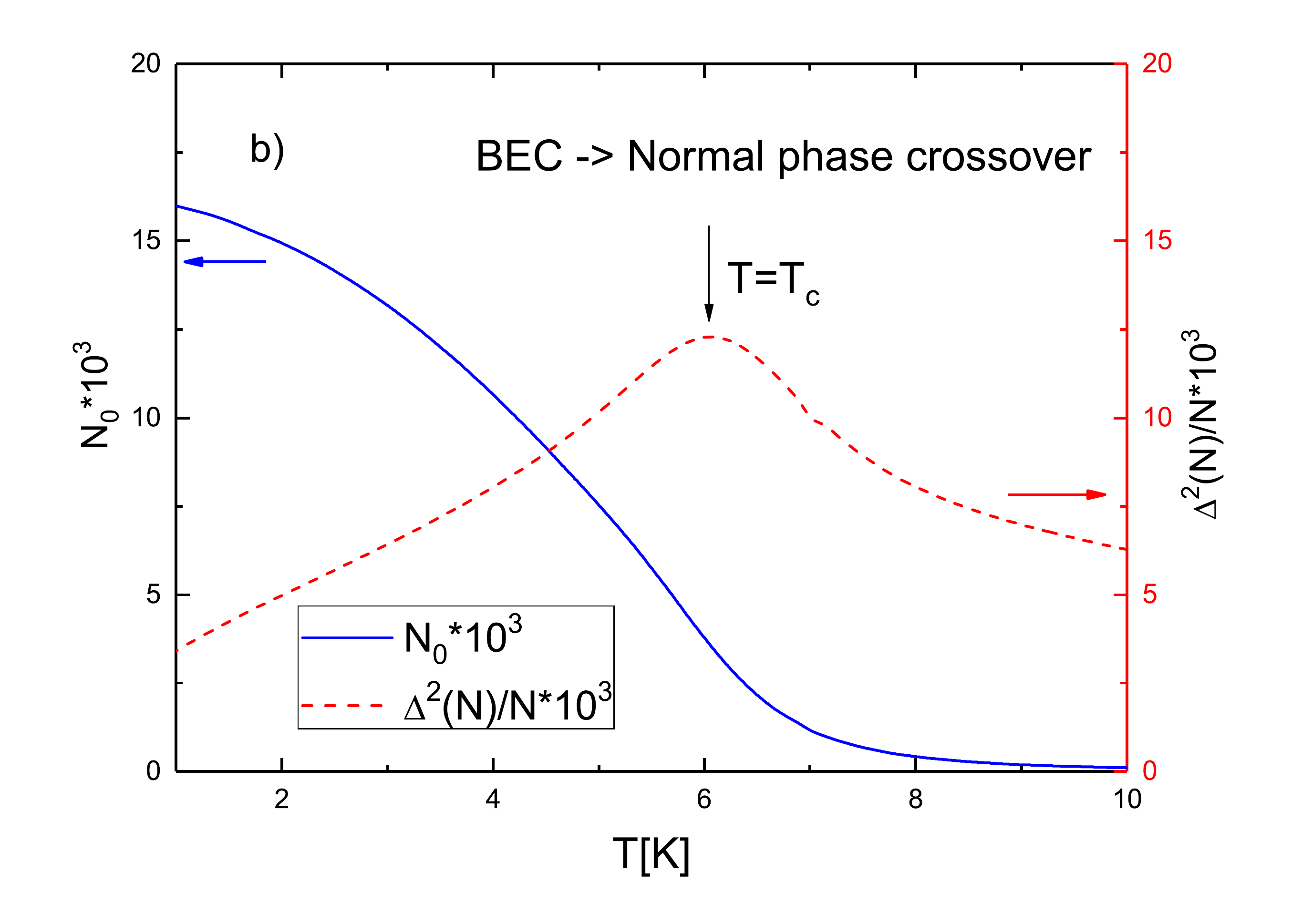}}
\hfill
\end{minipage}
\caption{The number fluctuations (dashed lines)  and condensed fractions  (solid lines) in a SOPT (a) and
 a smeared phase transitions (b).}
\label{Fig1}
\end{figure}

It is well known that in nature there exist not only second order phase transitions (SOPT) but also crossovers.
The former is characterized by its order parameter which goes to zero exactly at a certain critical temperature $T=T_c$. 
 This is illustrated in  Fig.\ref{Fig1}a (solid line), where $T_c$ is always interpreted as a critical temperature.
 The latter also has its own ``order parameter", which diminishes asymptotically at high temperatures
 (see Fig.\ref{Fig1}b, solid curve). Such kind of phase transitions are observed in spin-gapped quantum magnets,
 whose magnetization versus temperature can be explained by BEC to normal phase transition of triplons (magnons)
 \cite{Zapf}.

In this work, we address the question of determining the critical temperature in a crossover from the
Bose-Einstein condensed phase to the normal phase.
To illustrate our proposal on  the characterization of the crossover, we study the temperature dependence of magnetization of spin-gapped quantum magnets described by BEC of triplons. In this context, triplons are bosonic quasi-particles that describe the singlet-triplet excitations by external magnetic field in spin-gapped magnetic materials. As there are singlet-triplet excitations, they are referred as ``triplons'' instead of magnons \cite{Zapf}.

The triplon gas  has its own specifications being compared with  an atomic gas \cite{ourctan2}. Particularly, 
 in tasks related to atomic Bose gases the number of particles $N$ is assumed to be fixed, while
the chemical potential $\mu(N,T)$ is to be calculated e.g., by the relation
$N\sim\sum_{k}/[e^{\beta(\veps_k-\mu)}-1]$, where $\beta$ is the inverse temperature\footnote{Here and below we adopt the units $k_B=1$ for the Boltzmann constant, $\hbar=1$ for the Planck constant, and V=1 for the unit cell volume.}.
As to the triplon gas, the chemical potential  characterizes an additional direct contribution to the
triplon energy due to the external magnetic field $H$, giving $\mu=g\mu_B(H-H_c)$ where $g$  is the electron
Land{\'e} factor,
 $\mu_B=0.672$ K/T is the Bohr magneton and $H_c$ is the critical magnetic field which defines the gap
 $\Delta_{ST}=g\mu_BH_c$ between singlet and triplet states.
In the field induced BEC, $\mu$ is assumed to be an input parameter, from which the total number of triplons
can be calculated.
Besides, for homogenous atomic gases one may use simple quadratic bare dispersion $\veps_k=k^2/2m$ with a good accuracy, while for spin-gapped quantum magnets a more complicated form of bare dispersion is needed \cite{ouraniz1}. 
  Moreover, it has been established that in some magnetic compounds such as Ba$_3$Cr$_2$O$_8$ with a good isotropic symmetry, the phase transition from BEC into a normal phase of triplons is of a second order \cite{aczel,ourmce}, while the existence of anisotropies
for example in TlCuCl$_3$ smears out the transition into a crossover \ci{Zapf,Sirker1,ouraniz2part1,ouraniz2part2,tanaka}.
In contrast to trapped atomic gases, the fraction of particles, $N_0$, in the condensate
can easily be measured as  $N_0\sim {M_{stag}^2} $, and the number of triplons
is $N\sim M$, where $M_{stag}$
and $M$ are the staggered and total magnetizations, respectively, caused by the external magnetic filed $H$ \ci{Sirker1}, which defines the chemical potential. 

The first problem we address is how to define an effective critical temperature for such crossover.
In other words, it is curious to know if there is a physical quantity, which has at least a local extremum near a specific temperature ($T_c$) both for the second order phase transitions and crossover. We will show that the answer is positive: it is the particle number fluctuations versus temperature. For the SOPT this is
illustrated in Fig.\ref{Fig1}a (dashed line).

The second problem which we discuss in this work is related to the scaling of fluctuations with atom number 
which can be classified either as normal or anomalous \ci{wilhelm, yukfluc}.

The rest of this paper is organized as follows. In Section 2 we outline the definition of fluctuations and present 
some of their general properties. In Section 3 we  outline our main methodology used for evaluation of 
 particle number fluctuations and related physical quantities for the system of triplons in spin-gapped magnets and present our results in Section 4. The discussions of results and conclusion will be presented in the last section. The explicit derivation of the main equations are presented in the Appendix.

\section{Normal and anomalous fluctuations}

For the convenience of a reader we start with the main definitions and properties of fluctuations. 
A fluctuation or a dispersion of any physical quantity is a measure of deviation from its average
value derived from many identical random processes.
Fluctuations are important to our understanding of classical and quantum systems \ci{patash}. They are ubiquitous in physics:
from the primordial quantum fluctuations in the early universe that reveal themselves as fluctuations in the cosmic microwave background, to current fluctuations in every-day conductors. Even the Casimir effect is caused by fluctuations (of vacuum), where the average electromagnetic field is zero:  $\langle{\hat{\bf E}}\rangle=\langle{\hat {\bf H}}\rangle=0$.

One may differentiate between normal and anomalous fluctuations. Let an observable quantity $A$ is given
by statistical average $\langle \hat{A}\rangle$ of a Hermitian operator $\hat{A}$. Its fluctuation
is characterized by the dispersion:
\be
\Delta^2(\hat{A})= \langle\hat{A}^2\rangle- \langle \hat{A}\rangle^2\, 
\ee
Further,  let $\Delta^2(\hat{A})\sim N ^{1+\alpha}$, where $N$ is the number of particles.
Fluctuation is referred to as  normal, if $\alpha=0$  and anomalous otherwise,  $\alpha \neq 0$. It can be shown that a system with an anomalous fluctuation is unstable. In fact, one may write the general form of the necessary stability condition as \cite{yukfluc}
 \be \label{stab}
 0<\frac{\Delta^{2}(\hat{A})}{N}<\infty
 \ee
 This condition is required for any stable equilibrium systems at finite temperature and for all $N$, including the limit $N\rightarrow\infty$. The value of (\ref{stab}) can be zero only at $T=0$. Additionally, one can see that while $\Delta^{2}(\hat{A})$ has the order of $N$, the dispersion is normal and stability condition is maintained. But when $\Delta^{2}(\hat{A})\sim N^{\alpha}$, with $\alpha>1$, then such a dispersion is called anomalous and the fluctuation of $\hat{A}$ is anomalous, since equation (\ref{stab}) becomes proportional to $N^{\alpha-1}$ and goes to infinity as $N\rightarrow\infty$. As a result, the stability condition is broken and anomalous fluctuations are unstable.

Using Maxwell  relations, one can express a fluctuation of a physical quantities in terms of thermodynamic observables \cite{yuktutorial}.
Below we provide some examples.

(i) Energy fluctuations:
\be
\frac{\Delta^2 (\hat H)}{N}=C_VT^2
\ee
where $\hat H$  is Hamiltonian operator and  $C_V$ is the heat capacity (at constant volume).

(ii) Particle number fluctuations:
\be
\frac{\Delta^2 (\hat N)}{N}=\kappa_T \rho T
\ee
where
$\kappa_T$ is the isothermal compressibility and  $\rho$ is the density of particles.

(iii) Magnetization  fluctuations:
 \be \label{flucM}
\frac{\Delta^2 (\hat M)}{N}=\frac{\langle {\hat M}^2\rangle - \langle \hat M\rangle^2}{N}=\chi T
\ee
in which  $\chi $  is the magnetic susceptibility  $\chi=dM/dH$.
Note that all of these fluctuations are normal and these expressions are relevant to any thermodynamically stable system in the equilibrium \ci{yuksyms}.

\subsection{Anomalous fluctuation in BEC}

However, it is well known that, near the point of phase transition the system become unstable.  A phase transition occurs
exactly because one phase becomes unstable and has to change to another stable phase and further,
after a phase transition has occurred, the system  becomes stable again
and susceptibilities become finite.. We know that exactly at the critical temperature of a second order phase transition the fluctuations become anomalous \ci{yuksyms}. However, after the phase transition  has occured, the system becomes stable again. In Fig.\,1a (dashed line) we present the example of number  fluctuations.
As it is seen from Fig.\ref{Fig1}a (solid line), at $T=T_c$ the condensed fraction $N_0$  vanishes, but the fluctuation in number of particles becomes very large. Moreover, at exactly
$T=T_c$ the fluctuations become anomalous. It is interesting to explore whether the fluctuations
remain anomalous below the critical temperature also. That is, whether $\alpha(T<T_c ) \neq 0$  or not.
Unfortunately, in the literature it has not still been established if the fluctuations in condensed fraction
are normal or anomalous. Actually, on the one hand, Giorgini et al. \ci{giorgini} and Christensen et al. \ci{Christensen}
found for the fluctuations in condensed fraction $\alpha\approx 0.134$, which means that the fluctuations remain anomalous even in the whole BEC phase.  Thus, in many works on Bose systems, it is argued that
fluctuations remain anomalous far below the condensation point, in the whole
region of the Bose-condensed system.

On the other hand, Yukalov \ci{yuksyms,yukfluc} has proven that there are no anomalous fluctuations
in stable equilibrium systems.  Briefly, he rigorously proved the following theorem.
The dispersion of a global observable is normal if and only if all partial dispersions of its terms are normal, and it is
anomalous if and only if at least one of the partial dispersions is anomalous. In other words,
if ${\hat A}={\hat A}_1 + {\hat A}_2+....+{\hat A}_n $ and ${\hat A}$  is normal, then each ${\hat A}_i $  should be normal
and vice versa. Particularly, for a BEC system we have ${\hat N} = {\hat N}_0  +{\hat N}_1$, with
$N_0$ the number of condensed particles. Since ${\hat N}$ is normal i.e.  has a normal fluctuation, then ${\hat N}_0$ as well as ${\hat N}_1$  should be normal. In the next section we shall consider the consequences of this statement on the example of triplon BEC.

\section{Magnetization fluctuations in BEC of triplons in spin-gapped magnets}

Presently, it has been established that \ci{pitbook, yukquas} not only atoms but also quasiparticles may undergo BEC.  The experiments on magnetization of
spin-gapped magnets, (see the review by  Zapf et al. \ci{Zapf})  can be explained by BEC of triplons,
with $ M=g\mu_B N$. At low temperatures, $(T\leq T_c)$ triplons are condensed, which  leads to increase of magnetization of the antiferromagnets, e.g., TlCuCl$_3$.
Moreover, the experiments on staggered magnetization $M_{stag}$ by Tanaka et al. \ci{tanaka} have revealed that
staggered magnetization and hence the condensed fraction in TlCuCl$_3$ diminishes smoothly, not abruptly as in the case of pure BEC.  This means that the condensate density
is  non-zero for all temperatures due to the relation $M_{stag}^2=(g\mu_B)^2 N_0$. Thus, one may conclude that the  phase transition of triplons from BEC
into the normal phase, in general, is not of second order. In fact, it is a smeared phase transition, so that there is no
fixed temperature where the order parameter exactly would be  equal to zero.

Below we briefly outline the methodology we used to evaluate magnetizations and number fluctuations.

\subsection{Methodology}
It has been shown that \ci{Sirker1, ouraniz2part1}, the reason of this phenomena is the existence of exchange (EA) and Dzyaloshinsky-Moriya (DM) anisotropies. The effective Hamiltonian of a triplon gas can be presented as the sum of ``isotropic'' and ``anisotropic'' terms \ci{ ouraniz2part1, ouraniz2part2}
\begin{subequations}
\begin{align}
&{\cal H}=H_{iso}+H_{aniso}, \label{eq:H}\\
&H_{iso}=\int d\bfr\left[\psi^\ast(\bfr) (\hat{K}-\mu) \psi(\bfr)+ \frac{U}{2} (\psi^\ast(\bfr)\psi(\bfr))^2\right], \label{eq:H1}\\
&H_{aniso}=H_{EA}+ H_{DM},\label{eq:H2}\\
&H_{EA}=\frac{\gamma}{2} \int d\bfr \left[\psi^\ast(\bfr)\psi^\ast(\bfr) + \psi(\bfr)\psi(\bfr) \right], \label{eq:H3} \\
&H_{DM}= i\gamma'\int d\bfr \left[\psi(\bfr)-\psi^\ast(\bfr) \right],
 \label{eq:H4}
\end{align}
\label{eq:Htot}
\end{subequations}
where $\psi(\bfr)$ is the bosonic field operator, $U, \gamma, \gamma'$ are the interaction strengths
($U\geq 0, \gamma \geq 0, \gamma' \geq 0$) and $\hat{K}$ is the kinetic energy operator which defines the bare triplon dispersion $\varepsilon_k$ in momentum space. The integration is performed over the unit cell of the crystal with corresponding momenta defined in the first Brillouin zone \cite{ouraniz1}.  The linear Hamiltonian, an external source, in Eq.\,\eqref{eq:H4} corresponds to a simple case when singlet-triplet mixing is neglected and DM vector is chosen as $D \parallel x$ and $H\parallel z$ \cite{Sirker1}.   Thus, once the Hamiltonian is given, one first separates fluctuations as
$\psi=\xi\sqrt{\rho_0}+\tilde{\psi}$, where $\xi=e^{i\Theta}$ and $\rho_0$ are the phase of the condensate wave
function and its magnitude, respectively; and then introducing second quantization, $\tilde{\psi}=\sum_ke^{i\bfk\bfr}a_k$,  $\tilde{\psi}^\ast=\sum_ke^{-i\bfk\bfr}a_k^\ast$, makes an attempt to diagonalize the Hamiltonian $H$ with respect to creation ($a^\ast$) and annihilation ($a$) operators. As a result, analytical expressions for quasiparticle (bogolon) dispersion $E_k$ and some other quantities may be obtained. In the present work, we shall take into account anomalous averages $\sigma=\sum_{k}\sigma_{k}=\frac{1}{2}\sum_k
     \left(\langle a_{k}a_{-k}\rangle+\langle a_{k}^{\dag}a_{-k}^{\dag}\rangle\right) $ ($\sigma$-anomalous density) based on Hartree-Fock-Bogoliubov approach, which was neglected in  \cite{Sirker1}. This allows one to obtain continuous magnetization across the BEC transition, which would be discontinuous otherwise, in the so-called Hartree-Fock-Popov (HFP) approximation with $\sigma=0$ \ci{ourANN}.

In order to get more information about thermodynamics of the system we exploit the grand canonical thermodynamic
potential $\Omega$, which may be evaluated in the path integral formalism \cite{cooper,andersen,ouryee,klbookfi}. This will be convenient to study the modification of the
condensate wave function, entropy $S=-(\partial\Omega/\partial T)$, heat capacity $C_H=T(\partial S/\partial T)$, magnetization $M=-(\partial\Omega/\partial H)$ as well as magnetization fluctuations given by Eq.\,\eqref{flucM}. For the convenience of a reader the detailed calculations and the explicit expressions for these quantities are moved to the Appendix.

\section{Results}

Clearly, to obtain realistic numerical results one should find optimal input parameters of the Hamiltonian. 
The effective Hamiltonian in \eqref{eq:Htot} has mainly the following input parameters: $U, \gamma$ and $\gamma'$. Unfortunately, their optimal values have been fairly known in the literature \footnote{The set of parameters, proposed by Sirker et.al. \ci{Sirker1} could not be reliable, since in their HFP approximation they didn't take into account the anomalous density  $\sigma$}. There, using the method of least squares in data fitting we have found optimal values as $U=367$K, $\gamma=0.05$K and $\gamma'=0.001$K, corresponding to the best   description of experimental data on magnetizations  \ci{tanaka, tanaka1,recenttan} of TlCuCl$_3$  as it is shown in Figs. 2. This set of parameters can be considered as one of our main results, since it may be used in theoretical description of spin-gapped magnets such as TlCuCl$_3$ and KCuCl$_3$.

\begin{figure}[h]%
\centering
\begin{minipage}[H]{0.45\linewidth}
\center{\includegraphics[width=1.25\textwidth]{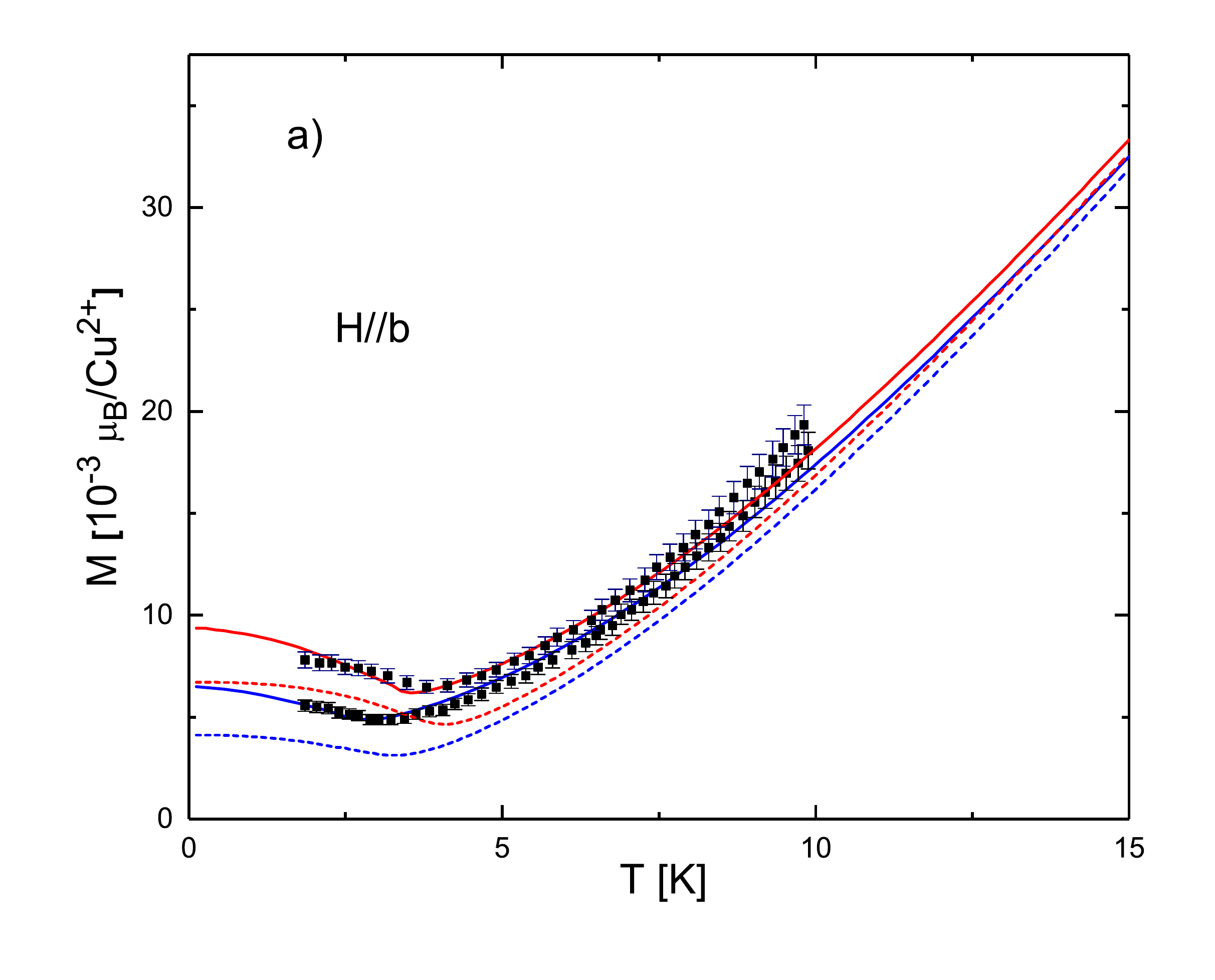}}
\end{minipage}
\hfill
\begin{minipage}[H]{0.5\linewidth}
\center{\includegraphics[width=1.25\textwidth]{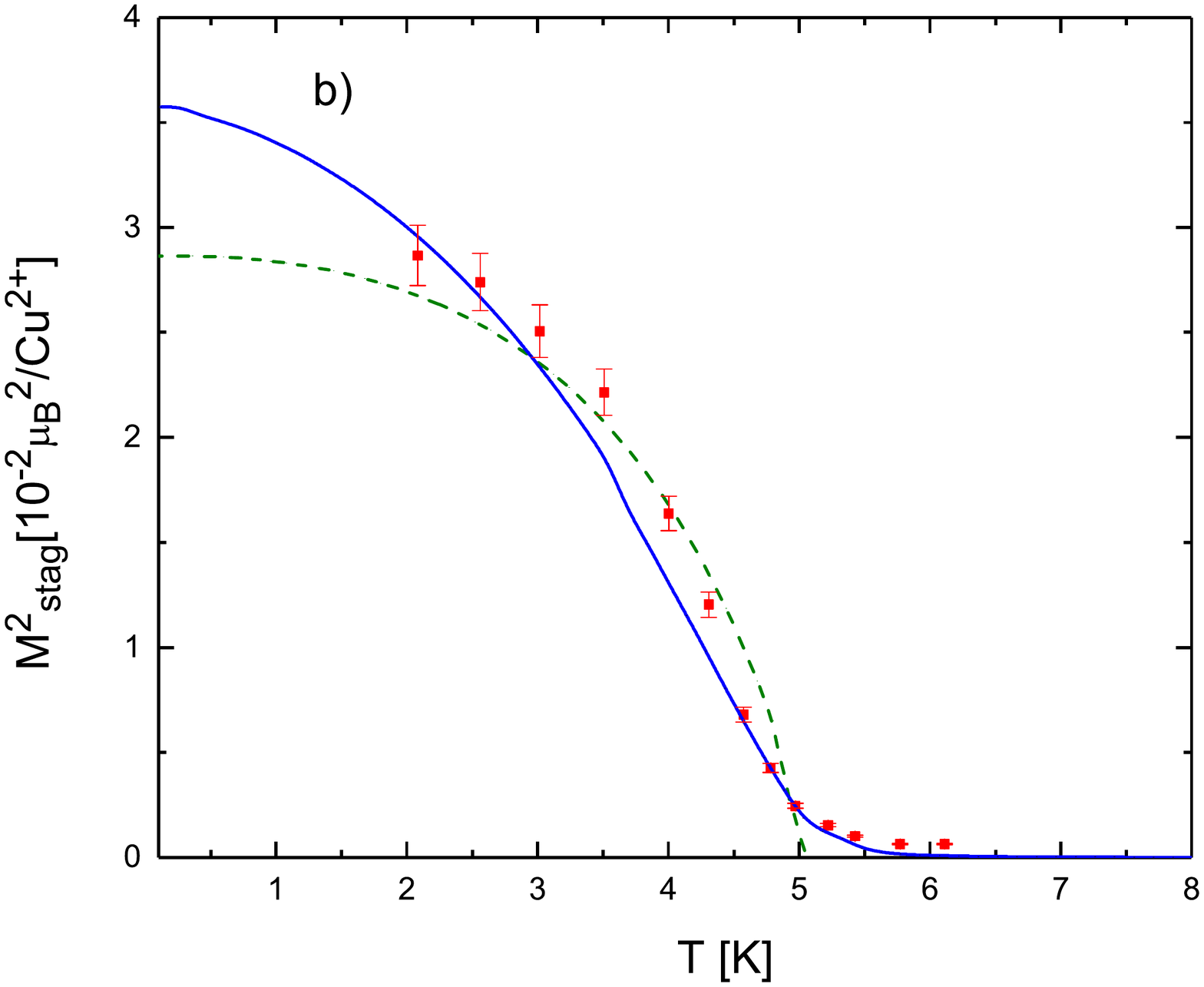}}
\end{minipage}
\caption {Uniform (a) and staggered (b) magnetizations for TlCuCl$_3$,  $H // b$.
Solid and dashed lines correspond to the present approximation and approximation in Ref.\cite{Sirker1}, respectively. Experimental data are taken from Ref.\cite{tanaka,recenttan}.
Following set of input parameters are used: $U=367$K, $\gamma=0.05$K and $\gamma'=0.001$K}
\label{Fig2}
\end{figure}

\begin{figure}[h]%
\centering
\begin{minipage}[H]{0.45\linewidth}
\center{\includegraphics[width=1.25\textwidth]{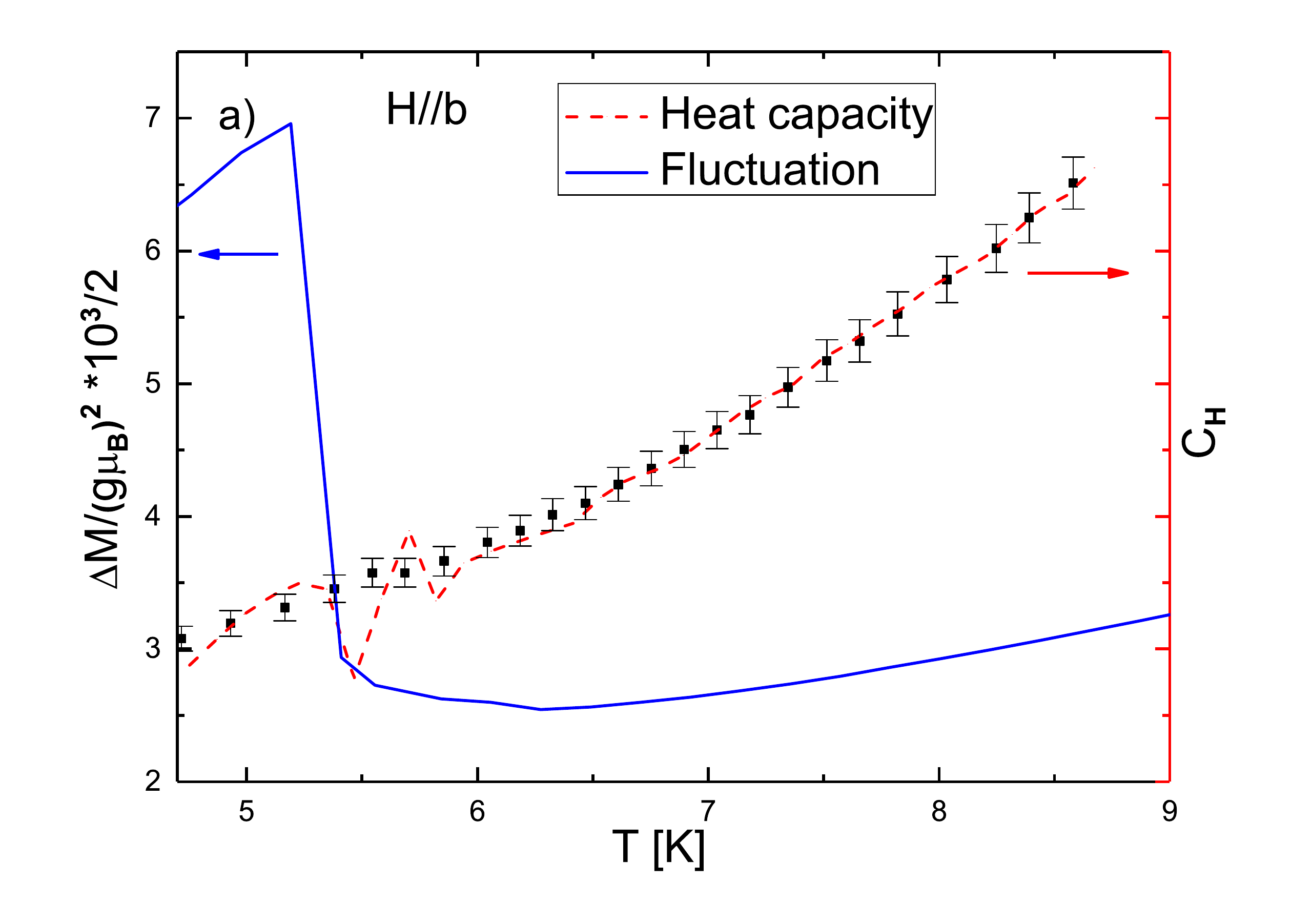}}
\end{minipage}
\hfill
\begin{minipage}[H]{0.45\linewidth}
\center{\includegraphics[width=1.25\textwidth]{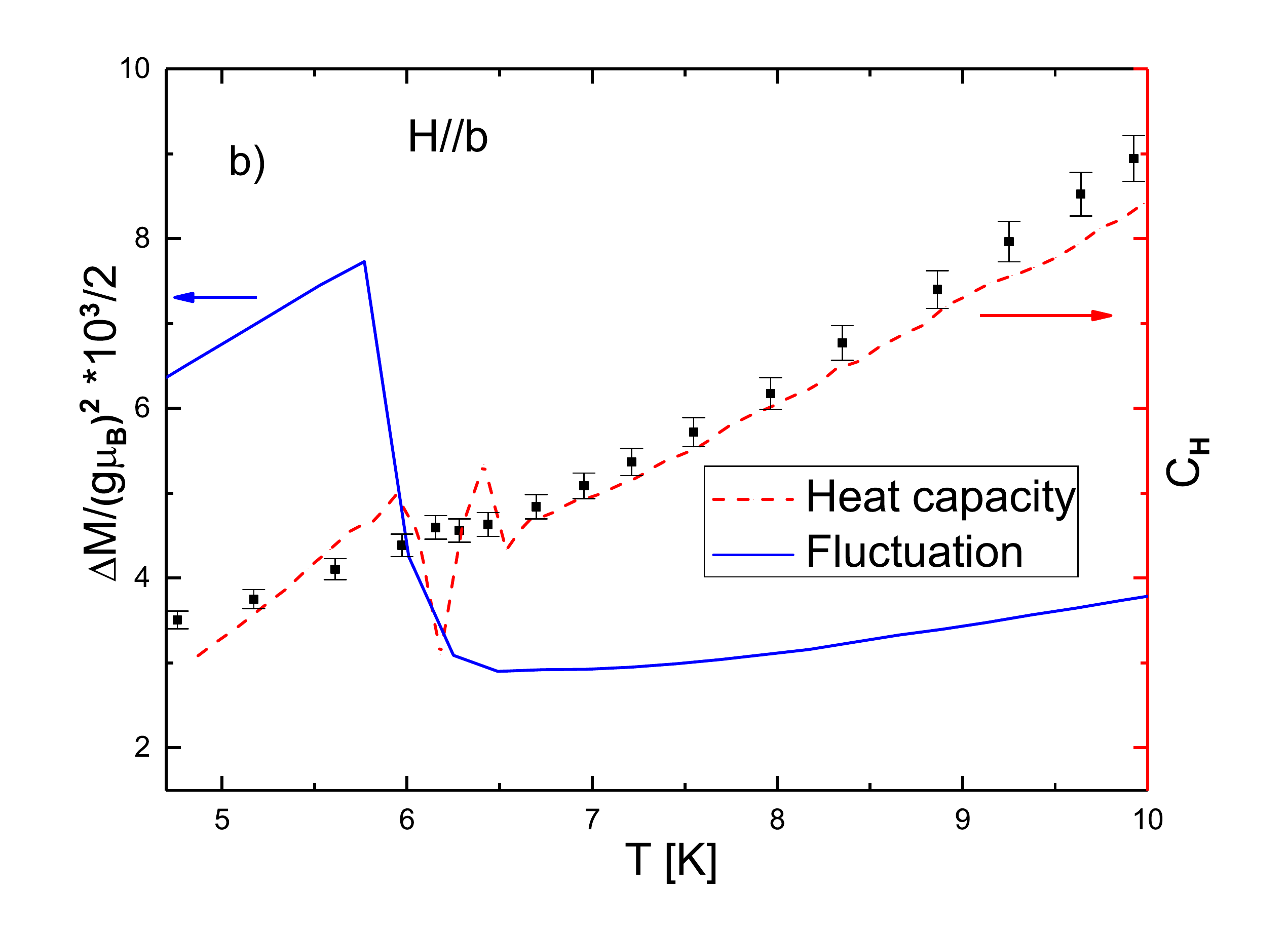}}
\end{minipage}
\caption {The fluctuations in magnetization (solid lines) and the heat capacity (dashed lines) for
TlCuCl$_3$ at $H=9$\,T (a) and $H=10$\,T  (b), respectively. It is seen that, in the region of phase transition
the heat capacity is smeared, but the fluctuation has a sharp maximum at a definite temperature, which may be interpreted as
a critical temperature of such crossover.}
\label{Fig3}
\end{figure}

Now, we are on the stage of studying magnetic fluctuations and heat capacity at constant external magnetic field ($C_H$) in TlCuCl$_3$. In Fig.\ref{Fig3} we present the fluctuations in magnetization (solid lines) and heat capacity (dashed lines) for TlCuCl$_3$  at $H=9$ T (Fig.\ref{Fig3}a) and $H=10$ (Fig.\ref{Fig3}b). It is seen that near phase transition the heat capacity is rather smeared, while the fluctuations in magnetizations has a sharp maximum. Therefore, e.g. $T\simeq5.4$K may be considered as a critical temperature of this crossover for $H=9$T. It is also seen that $\Delta M$  increases at high temperatures as it is expected from  the Eq.\,\eqref{flucM}. On the other hand, it  should be noted that the fluctuation vanishes at exactly $T=0$, owing  to the relations $\Delta \hat{M}\sim \Delta \hat{N}$ and $\Delta\hat{N}\mid_{T=0}=0$ \cite{yuksyms}.

In some cases, a crossover has its precursor effect. For example, BCS-BEC crossover includes a pseudogap region \cite{nature}. However, there are no precursor effects in spin-gapped quantum magnets under consideration. Note that, our general assumption on the relation between the critical temperature and maximum of fluctuations is in good agreement with pair fluctuation approximation \cite{prb66} where the critical temperature corresponds to the pole of pair fluctuation parameter.

\section{Discussions and Conclusion}
In this work, we looked at the problem of defining a critical temperature in a crossover from the BEC phase to the normal phase by studying the temperature dependence of magnetization of spin-gapped quantum magnets described by BEC of triplons. We have calculated the heat capacity $C_H$ at constant field and fluctuations in magnetization  using the Hartree-Fock-Bogouliubov approximation and found optimized parameters of the Hamiltonian of triplon gas. In the region of phase transition, the heat capacity $C_H$ is
smeared out due to the Dzyaloshinsky-Moriya interaction. The sharp maximum
of the fluctuations in the magnetization is identified as the critical temperature of the crossover.

We found that there is no anomalous fluctuation in the condensed fraction. This is simply because in the mean-field
approximation one always uses Bogoliubov shift
$ {\psi}= \sqrt{\rho_0}+ { {\tilde{\psi}}}$
where $ {\psi}$ and ${\tilde{\psi}}$ are operators, however,
the first term here is not an operator, but  just a function. Thus, from the definition of  the dispersion  one gets
$ \Delta ^2 (\rho_0)=\langle\rho_{0}^{2}\rangle - \langle\rho_{0}\rangle^2 =0$.
We have also found optimized values of coupling constants, $U$, $\gamma$ and $\gamma'$ for  TlCuCl$_3$,
which are the strengths of the contact triplon-triplon interaction, exchange anisotropy and DM interactions,
respectively. These parameters  may be used to further study of the physical properties of this material.

We have shown that, for the crossover  phase transition the heat capacity is smeared, while, the number fluctuation, which is proportional to the magnetization has a maximum. This point may be interpreted as a critical temperature of the crossover.
It would be interesting to study the number (or density) fluctuations for other smeared phase transitions such as 
BCS-BEC  \ci{Burovski,Klimin} and magnetic transitions in sodium rich materials, e.g., Na$_x$CoO$_2$  \ci{baran}. 
It is beyond the scope of the present paper to have a detailed discussion of BCS to BEC, thus the reader is 
referred to recent lectures by Zwerger \cite{wilhelm2}.
There are also smeared phase transitions due to disorder in binary alloys. Such materials consist of
substances A and B. Material A is in the magnetic phase while material B is in the paramagnetic phase.
The smeared phase transition, as well as quantum ones,  from a nonmagnetic to a magnetic phase can be
tuned by substituting magnetic atoms A for nonmagnetic atoms B in the binary alloy
A$_{1-x}$B$_{x}$ \ci{nozadze}. It  would be interesting to study fluctuations in such materials, to see if
they have a maximum at some value of $x$ or temperature.
 
\acknowledgements
We are indebted to participants of the conference ICSM-21 for useful discussions and comments.
This work is supported by the Ministry of Innovative Development of the Republic of
Uzbekistan and the Scientific and the Technological Research Council of Turkey (TUBITAK)
under Grant No.\,119N689.

\section*{Appendix: Derivation of thermodynamic quantities}
\def\theequation{A.\arabic{equation}}
\setcounter{equation}{0}

\bea
\psi(\mathbf{r},t)= i\sqrt{\rho_0} + \ti{\psi}(\mathbf{r},t), \quad  \psi^\dagger(\mathbf{r},t)= -i\sqrt{\rho_0}  + \ti{\psi}^\dagger(\mathbf{r},t)
\label{eq:psi}
\eea
where $\rho_0$ is density of condensed particles and $\ti{\psi}$ is fiedd operator of uncondensed particles.

One of our main goals is to find an analytical expression for thermodynamic potential $\Omega$, which contains almost all the information about the equilibrium statistical system  \cite{yuktutorial}. For this purpose we use the path integral formalism where $\Omega$ is given by
\begin{equation}	
\Omega = - T\ln{Z}, \qquad Z  = \int  D \ti\psi^\dagger D \ti\psi e^{-A[\psi,\psi^\dagger]}.
\label{eq:entropy}
\end{equation}
Here, $A[\psi,\psi^\dagger]$ is the action and its form can be chosen by the total Hamiltonian (\ref{eq:Htot}) as follows
\begin{eqnarray}
&A[\psi,\psi^\dagger]= \nn
&\int_{0}^{\beta}d{\tau} d\mathbf{r}\left( \psi^\dagger K_{id} \psi + \frac{U}{2}\left( \psi^\dagger\psi\right)^2  +\frac{\gamma}{2}\left(\psi^\dagger\psi^\dagger+ \psi\psi\right) +i \gamma'\left( \psi-\psi^\dagger\right)\right)
\label{eq:entropy1}
\end{eqnarray}
where $K_{id}=\frac{\partial}{\partial\tau}-\mathbf{\hat{K}}-\mu$. For simplicity, we dropped arguments of wave function operators. In Eq.\,\eqref{eq:entropy1} the fluctuating fields $\ti{\psi}(\bfr,\tau)$ and $\ti{\psi}^\dagger(\bfr,\tau)$ satisfy the bosonic commutation relations and periodic in $\tau$ with period $\beta=1/T$. Clearly, this path integral can not be evaluated exactly, so an approximation is needed. In the present work, we  used an approach, which is called the variational perturbation theory \cite{stancu}, or $\delta$-expansion method.
To apply this method,  we maked following replacements in the action (\ref{eq:entropy1}): $U\rightarrow \delta U$, $\gamma\rightarrow \delta \gamma$, $\gamma'\rightarrow \sqrt{\del}\gamma' $. And added to the action the term
  \bea
   A_\Sigma= (1-\delta)\int d{\tau} d\mathbf{r}\left[ \Sigma_n\ti{\psi}^\dagger\ti{\psi} + \frac{1}{2} \Sigma_{an}\left(\ti{\psi}^\dagger\ti{\psi}^\dagger+\ti{\psi}\ti{\psi} \right) \right]
   \label{eq:Asigma}
  \eea
where the variational parameters $\Sigma_n$ and $\Sigma_{an}$ may be interpreted as the normal and anomalous
self-energies, respectively. They are defined as \cite{andersen}:
 \begin{subequations}
	\begin{align}
&\Sigma_{n}=(\Pi_{11}(0,0)+\Pi_{22}(0,0))/2,\\ &\Sigma_{an}=(\Pi_{11}(0,0)-\Pi_{22}(0,0))/2, \\ &\Pi_{ab}(\omega_{n},\mathbf{k})=(G(\omega_{n},\mathbf{k}))^{-1}_{ab}-(G^{0}(\omega_{n},\mathbf{k}))^{-1}_{ab}
\label{eq:9}
 \end{align}
\end{subequations}
and the Green functions $G(\omega_n, \mathbf{k})$, $G^{0}(\omega_n, \mathbf{k})$ are given further below.
Now, we insert (\ref{eq:psi}) into the action (\ref{eq:entropy1}) and divide this action into five parts according to orders of $\tilde{\psi}$ (see \cite{ouraniz2part1} for more detailed calculations)
\be
A=A_0+A_1+A_2+A_3+A_4.
\ee
Then, writing $\ti{\psi} $, $\ti{\psi}^\dagger$ in Cartesian form as
\begin{subequations}
	\begin{align}
 \ti{\psi}&= \frac{1}{\sqrt{2}}(\psi_1 + i \psi_2).\\
 \ti{\psi}^\dagger&= \frac{1}{\sqrt{2}}(\psi_1 - i \psi_2).
 \end{align}
\end{subequations}
the form of grand partition function $Z$ can be obtained in (\ref{eq:entropy}) and that helps us to calculate $\Omega$.
The perturbation scheme may be considered as an expansion in powers of $\delta$ by using the Greens functions
\bea
G_{ab}(\tau, \mathbf{r}; \tau', \mathbf{r}')=\frac{1}{\beta}\sum_{n,k} e^{i\omega_n(\tau-\tau')+i\mathbf{k}(\mathbf{r}-\mathbf{r}')} G_{ab}(\omega_n, \mathbf{k})
\label{eq:Green1}
\eea
$(a, b= 1,2)$, where $\omega_n=2\pi nT$ is the $n$th bosonic Matsubara frequency and
\bea
G_{ab}(\omega_n, \mathbf{k}) = \frac{1}{\omega_n^2 + E_k^2}\begin{bmatrix}
	\epsilon_k+X_2 & \omega_n \\
	-\omega_n & \epsilon_k+X_1
	
\end{bmatrix}.
\label{eq:Gab}
\eea
In Eq.\,\eqref{eq:Gab} $E_k$ corresponds to the dispersion of quasi-particles (Bogolons)
\bea
 E_k = \sqrt{\epsilon_k+X_1}\sqrt{\epsilon_k+X_2}.
 \label{eq:energy}
\eea
where is the bare dispersion
of triplons and the self-energies $X_1$ and $X_2$ which are given by
 \begin{subequations}
 \begin{align}
 X_1 =\Sigma_{n}+\Sigma_{an}-\mu, \label{eq: x1}\\
 X_2 = \Sigma_{n}-\Sigma_{an}-\mu.
 \label{eq: x2}
 \end{align}
\end{subequations}

Finally,  the following expression for $\Omega$ can be obtained including EA and DM interactions
\begin{subequations} \label{eq:omega}
	\begin{equation}
    \Omega= \Omega_{ISO} + \Omega_{EA} + \Omega_{DM}.
	\end{equation}
  \begin{align}
	\Omega_{ISO}= -\mu \rho_0 + \frac{U\rho_0^2}{2} +\frac{1}{2}\sum_k(E_k-\epsilon_k) +T \sum_k \ln (1-e^{-\beta E_k})\nn
	\qquad\qquad\qquad\qquad + \frac{1}{2}(\beta_1 B +\beta_2 A) +\frac{U}{8}(3A^2+3B^2+2AB), \label{eq:omega1}\\
   \Omega_{EA}= -\gamma \rho_0 +\frac{\gamma}{2}(B-A), \label{eq:omega2}	\\
     \Omega_{DM}  = -2\gamma'\sqrt{\rho_0} -\frac{\gamma'^2}{X_2}. \label{eq:omega3}
	\end{align}
\end{subequations}
where
 \begin{subequations}
 	\begin{align}
 	\beta_1& =-\mu -X_1 + U\rho_0\\
 	\beta_2& = - \mu -X_2 + 3U\rho_0\\
 	A&= T\sum_{k,n} \frac{\epsilon_k+X_1}{\omega_n^2+E_k^2}= \sum_k W_k \frac{\epsilon_k+X_1}{E_k}\\
 	B&= T\sum_{k,n} \frac{\epsilon_k+X_2}{\omega_n^2+E_k^2}= \sum_k W_k \frac{\epsilon_k+X_2}{E_k}
 	\end{align}
 \end{subequations}
in which $W_k =1/2 +1/(e^{\beta E_k}-1)$ and  $$\sum_{n, \mathbf{k}}\equiv\sum_{n=-\infty}^{n=\infty}\int d\mathbf{k}/(2\pi)^3 $$.

The stability condition in the equilibrium requires, $X_1 \geq 0$, $X_2 \geq 0$ as well as for the grand thermodynamic potential satisfies that
 \bea
 \frac{\partial \Omega}{\partial X_1}=0, \quad  \frac{\partial \Omega}{\partial X_2}= 0, \quad \frac{\partial \Omega}{\partial \rho_0}= 0.
 \label{eq: partial}
 \eea
This condition leads to the following equations for $X_1$ $X_2$ and $\rho_0$
\begin{subequations}
	\begin{align}
X_1&= 2U\rho + U\sigma -\mu - U\rho_0+\gamma +\frac{2\gamma'^2 D_1}{X_2^2} \label{eq:X1}\\
X_2&=2U\rho - U\sigma -\mu +U\rho_0-\gamma -\frac{2\gamma'^2 D_2}{X_2^2} \label{eq:x2}\\
\mu&=U (\rho_0 +2\rho_1)-U\sigma - \gamma -\frac{\gamma'}{\sqrt{\rho_0}}
\label{eq:mu}
\end{align}
\end{subequations}
where
\begin{subequations}
	\begin{align}
	A_1'&= \frac{\partial A}{\partial X_1}= \frac{1}{8}\sum_{k}\frac{(E_k W_k' + 4W_k)}{E_k}\\
    A_2'&= \frac{\partial A}{\partial X_2}=\frac{1}{8}\sum_{k}\frac{(\epsilon_k + X_1)^2(E_k W_k' - 4W_k)}{E_k^3}\\
    B_1'&= \frac{\partial B}{\partial X_1} = \frac{1}{8}\sum_{k}\frac{(\epsilon_k + X_2)^2(E_k W_k' - 4W_k)}{E_k^3}\\
   D_1&=\frac{A_1'}{\bar{D}}; \quad D_2=\frac{B_1'}{\bar{D}}; \quad \bar{D}=A_1'^2 - A_2'B_1'\\
    W_k'& =\beta (1-4W_k^2)=\frac{-\beta}{\sinh^2(\beta E_k/2)}.
    \label{eq:parameter}
	\end{align}
\end{subequations}
Now, solving these (\ref{eq:X1}), (\ref{eq:x2}) and (\ref{eq:mu}) with respect to $X_1$, $X_2$ and $\rho_0$ one can find other physical quantities. Additionally, the crtical temperature $T_c$ is also found with these equations. However, application of these equations  for  $T>T_c$ and $T<T_c$ regions  should be written separately \cite{ouraniz2part2}. For a homogeneous system the normal and anomalous densities are defined as $\rho_1 = \int \langle\ti{\psi}^\dagger(r)\ti{\psi}(r)\rangle d\mathbf{r}$ and $\sigma=\int \langle\ti{\psi}(r)\ti{\psi}(r)\rangle d\mathbf{r}$ respectively, and may be calculated using the Green functions given in Eqs.\,\eqref{eq:Green1} and \eqref{eq:Gab}. As a result, they take the following explicit form
\begin{subequations}
	\begin{align}
	\rho_1 &= \frac{A+B}{2} = \sum_k\left[\frac{W_k(\epsilon_k+X_1/2 +X_2/2)}{E_k} -\frac{1}{2} \right] \equiv\sum_k \rho_{1k}, \label{eq:17a}\\
	\sigma &= \frac{B-A}{2} =\frac{(X_2-X_1)}{2} \sum_k \frac{W_k}{E_k}  \equiv\sum_k \sigma_{k}.
	\label{eq:17b}
	\end{align}
\end{subequations}
The total density of triplons per dimer is the sum of condensed and uncondensed fractions:
 \bea
  \rho =\frac{N}{V}=\rho_0 +\rho_1.
 \eea
Now, total and staggered magnetizations can be calculated by
\bea
M=g\mu_B\rho,    \quad   M_{stag}=g\mu_B\sqrt{\rho_0}.
\eea


\end{document}